\documentclass[reprint, amsmath, amssymb, aps, superscriptaddress]{revtex4-1}
\usepackage{packages}
\allowdisplaybreaks

\begin{document}

\preprint{APS/123-QED}

\title{Anisotropic thermalization of dilute dipolar gases}

\author{Reuben R.W. Wang}
\author{John L. Bohn}
\affiliation{JILA, University of Colorado, Boulder, Colorado 80309, USA}

\date{\today} 

\begin{abstract}

We study collisional rethermalization in ultracold dipolar thermal gases, made intricate by their anisotropic differential cross sections. Theoretical methods are provided to derive the number of collisions per rethermalization \cite{Monroe93_PRL}, which for dipolar gases, is highly dependent on the dipole alignment axis. These methods are formulated to be easily applied in experimental contexts, even reducing to analytic expressions if the route to thermal equilibrium is governed by short-time dynamics. In the analytic case, collisional rethermalization is fully characterized by the dipole magnitude and orientation, scattering length, and thermalization geometry. These models compare favorably to Monte Carlo simulations, and are shown to model well a recent experimental result on the rethermalization of polar molecular samples.

\end{abstract}

\maketitle

\section{\label{sec:introduction} Introduction}

A fundamental property of a gas brought out of thermal equilibrium and then left alone, is that it will return to equilibrium.  For a classical gas, equilibration occurs at a rate set by collisions among the constituents of the gas.  It follows that if the collision cross sections are anisotropic, so too might be the relaxation rate. 

Ultracold gases whose constituent atoms or molecules possess either magnetic or electric dipole moments are ideal environments to study this anisotropy.  In a cold enough gas, the polarization of its constituents can be maintained after collision, thus the intrinsic anisotropy of collision cross sections, defined by the polarization axis, persists throughout the thermalization.  In the controlled environment of the trap, the gas can be brought out of thermal equilibrium along a given axis (for instance by parametric heating or a trap frequency quench), while the resulting temperature evolution can be read out along any other desired axis, by, for example, imaging an expanded gas after turning the trap off.  

Experiments of this kind have been used in the past to determine scattering cross sections of ultracold atoms \cite{Monroe93_PRL, Arndt97_PRL, Hopkins00_PRA, Weinstein02_PRA, Ravensbergen18_PRA}.  Generally, the relaxation rate is proportional to the total collision rate.  The dimensionless ratio of these rates, dubbed the ``number of collisions per rethermalization'' and here denoted ${\cal N}$,  gives information on the nature of the cross section itself, including its anisotropy. For $s$-wave collisions, this ratio has been determined to be ${\cal N}_s =2.5$, \cite{Roberts01_thesis, Schmidt03_PRL, Goldwin05_PRA}, and is in practice an experiment-independent quantity. The same value of ${\cal N}_s$ has been alternatively determined by spectroscopic methods (Ref.~\cite{Coslovsky17_PRA}, where ${\cal N}_s$ is referred to as ``collisional softness"). For $p$-wave scattering of identical fermions, the number of collisions per rethermalizaion has been measured to be ${\cal N}_p=4.1$ \cite{DeMarco99_PRL}. For dipolar scatterers however, the elastic scattering cross section is highly anisotropic, giving ${\cal N}$ its own anisotropy. Numerical simulations have verified this anisotropy in rethermalization of ultracold Er atoms \cite{Sykes15_PRA}, and were used to extract the scattering length of bosonic Dy \cite{Tang15_PRA}. 


In this paper we present theoretical methods to compute ${\cal N}$ via the method of averages. These are relevant for dilute gases that are ultracold yet not quantum degenerate, so their thermalization dynamics is governed by the Boltzmann equation. For a broad range of experimental conditions, the various functional forms of ${\cal N}$ are in fact analytic functions of scattering length, dipole length, and dipole orientation under suitable approximations. These formulas not only bring out explicitly the strong effects of anisotropy, but also should be useful in designing and interpreting experiments where ultracold atoms and molecules are brought out of thermal equilibrium.  Of note are the prospects to accelerate evaporative cooling with dipolar scattering \cite{Lu11_PRL, Lu12_PRL}, especially through the saddle-point method \cite{Hess86_PRB} where evaporation occurs only along a chosen axis.  


We describe our theoretical formulation of the system in Sec.~\ref{sec:formulation}, then proceed to its use for characterizing rethermalization in Sec.~\ref{sec:relaxation}. Specifics that arise when treating either systems of bosons or fermions are addressed in Sec.~\ref{sec:NCPR_bosons} and Sec.~\ref{sec:NCPR_fermions} respectively, along with examples which illustrate the appeals and shortcomings of our models. Conclusions are drawn in Sec.~\ref{sec:conclusion}.

\section{\label{sec:formulation} Formulation}

We consider a single species gas of $N$, ultracold but non-degenerate dipoles. The gas is harmonically confined in the trapping potential
\begin{align}
    U(\boldsymbol{q}) = \frac{1}{2} m \left( \omega_x^2 x^2 + \omega_y^2 y^2 + \omega_z^2 z^2 \right),
\end{align}
where $m$ is the particle mass and $\omega_{x, y, z}$ are the trapping frequencies associated to each spatial dimension. The gas is also taken to be sufficiently dilute and far from the critical temperature, such that Bose-enhancement and Fermi-blocking effects can be neglected. Furthermore, we limit our study to rarefied gases in which mean-field effects due to long-range dipolar interactions can be ignored.  In such a parameter regime, scattering length scales are completely defined by the $s$-wave scattering length $a$ (for bosons), and magnetic (electric) dipole moment $\mu$ ($d$), expressed as a dipole length $a_d = C_{\text{dd}}m / (8\pi\hbar^2)$ (where $C_{\text{dd}} = \mu_0\mu^2$ for magnetic dipoles and $C_{\text{dd}} = d^2 / \epsilon_0$ for electric dipoles, with $\mu_0$ and $\epsilon_0$ being the vacuum permeability and permittivity respectively). 

In a harmonic trap at equilibrium, the coordinates $q_i$ and momenta $p_i$ are described by Gaussian distributions. When mildly heated, it is conceivable that the distributions remain nearly Gaussian \cite{Krook1977}, whereby the gas' dynamics is described by the the mean values $\langle q_i^2 \rangle$, $\langle p_i^2 \rangle$, given by averages over phase space distribution $f(\boldsymbol{q}, \boldsymbol{p}, t)$:
\begin{align}
    \langle \ldots \rangle \equiv \frac{1}{N} \iint d^3p d^3 q \: f(\boldsymbol{q}, \boldsymbol{p}, t) \times ( \ldots ).
\end{align}
The near-equilibrium collective dynamics is well described by the Enskog equations of change \cite{Colussi15_NJP, Guerey99_PRA}:
\begin{subequations}
\label{eq:Enskog_eqns}
\begin{align}
    & \dfrac{d \langle q_j^2 \rangle}{d t} - \dfrac{2}{m}\langle q_j p_j \rangle = 0, \label{eq:Enskog_eqns_a} \\
    & \dfrac{d \langle p_j^2 \rangle}{d t} + 2m\omega_j^2\langle q_j p_j \rangle = \mathcal{C}[ p_j^2 ], \\ 
    & \dfrac{d \langle q_j p_j \rangle}{dt} - \dfrac{1}{m}\left\langle p_j^2 \right\rangle + m\omega_j^2 \langle q_j^2 \rangle = 0, \label{eq:Enskog_eqns_c}
\end{align}
\end{subequations}
where $j = x, y, z$ and ${\cal C}$ is the collision integral derived from the Boltzmann equation \cite{Reif}.
The collision integral, in turn, is expressed in terms of $f$ and the differential cross sections,
whereby this integral depends on the polarization axis of the dipoles.  For concreteness, we assume this axis lies in the $x$-$z$ plane and its direction is given by  $\hat{\boldsymbol{{\cal E}}} = (\sin\Theta, 0, \cos\Theta)$, whereby relaxation rates become functions of $\Theta$.

\section{\label{sec:relaxation} Anisotropic Relaxation}

Once the gas is brought out of equilibrium and left alone, it will tend toward equilibrium and finally arrive at a final temperature as required by the $H$-theorem \cite{Reif}. Yet, the {\it rate} at which equilibrium is achieved may depend on the axis that is probed in a given experiment.

These rates can be probed in a cross-dimensional rethermalization experiment. In such experiments, rethermalization occurs by means of elastic collisions that redistribute thermal energy throughout the gas at a rate proportional to the mean total cross section, ${\bar \sigma}$ \cite{Monroe93_PRL, Schmidt03_PRL, Newbury95_PRA, Tang15_PRA}. For anisotropic scatterers however, not every collision counts the same toward rethermalization: if the differential cross section favors forward-scattering, it is not useful, since the particles would effectively have the same momenta after collision as before. 

Generally, a gas that is heated along the $i$th coordinate and whose rethermalization is measured along the $j$th coordinate will rethermalize at a rate $\gamma_{ij}(\Theta)$. This rate depends on the density and temperature of the given experiment, whereby it is useful to compare the rethermalization rate to a standard collision rate $n {\bar \sigma} \langle v \rangle$,
where $n = (1/N) \int n(\boldsymbol{r})^2 d^3 r$ is the average number density of the gas, $\overline{\sigma}$ is the total elastic cross section averaged over incident relative momenta 
and $\langle v \rangle = \sqrt{16 k_B T_0 / (m \pi)}$ is the mean collision velocity. Thus for relaxation considerations, the rethermalization rate is proportional to the total collision rate,
\begin{align}
    \gamma_{ij}(\Theta) \equiv \frac{ n {\bar \sigma} \langle v \rangle }{ {\cal N}_{ij}(\Theta)} \label{eq:gamma_N_rln}
\end{align} 
whose proportionality constant ${\cal N}_{ij}$, is known as the number of collisions per rethermalization.  



In a harmonic trap, the position and momentum widths, quantified by $\langle q_j^2 \rangle$ and $\langle p_j^2 \rangle$, will experience out-of-phase oscillations at the trap frequency en route to rethermalization. This motivates the definition of a non-equilibrium pseudotemperature along each axis:
\begin{equation}
    \mathcal{T}_{j} = \frac{m \omega_{j}^{2} \langle {q_j}^{2} \rangle}{2 k_{B}} + \frac{\langle {p_{j}}^{2} \rangle}{2 m k_{B}}, \label{eq:pseudotemperatures}
\end{equation}
that suppresses such oscillations in time-evolution, easing the extraction of rethermalization rates in theoretical studies. Comparing this with experimental data should not be difficult, since common imaging techniques such as time-of-flight measurements also have oscillations suppressed by the explicit removal of a trapping potential. 

We now present two theoretical procedures to extract the rethermalization rates, and thus $\cal N$, from these out-of-equilibrium pseudotemperatures. These are: a) explicit, full solutions to Eqs.~(\ref{eq:Enskog_eqns}), from which a time constant is extracted from fits to ${\cal T}_j(t)$ and; b) an approximation to these solutions, extracted from the short-time decay rate behavior, which leads to analytical expressions for ${\cal N}$.

To evaluate the accuracy of these calculations, we utilize numerical particle simulations to generate pseudotemperature time traces. These are performed with direct simulation Monte Carlo (DSMC) methods similar to that in Ref.~\cite{Wang20_PRA}, which can themselves be used as a theoretical tool for the study of such systems \cite{Bird1970, Bonasera94_PRep, Wade11_PRA, Guerey99_PRA}. Briefly speaking, each simulation prepares the  particles' phase-space coordinates as sampled from Gaussian distributions. Albeit Gaussian, the gas can be initialized out-of-equilibrium by having its momentum and position space widths widened from thermal equilibrium along a chosen axis. Such techniques have been established to give good agreement with experimental data \cite{Tang16_PRL, Sykes15_PRA}.

\subsection{ \label{sec:1e_decay_rate} Anisotropic Relaxation: 1/e Decay Rate }

An operational definition of the rethermalization time $\tau$, is the $1/e$ decay time (i.e. the time taken for the pseudotemperature to reach $1/e \approx 0.368$ of the extent from its initial to its final values). This can be extracted with an exponential fit to data one would get from cross-dimensional rethermalization experiments. 

In the close-to-equilibrium scenarios envisioned here, the collision integrals required in the Enskog equations are linear functions of the observables, and can be solved analytically \cite{Wang20_PRA}\footnote{Higher phase-space density regimes may necessitate the inclusion of quantum statistical effects to the collision integral, as considered in Ref.~\cite{Vichi00_JLTP}. We ignore this here. }. To this end, we consolidate the Enskog variables into a nine-dimensional vector
\begin{align}
    \boldsymbol{\xi}(t) = \big[
        & m^2\omega_{z}^2\langle z^2 \rangle, \:\:
        \langle p_z^2 \rangle, \:\:
        m \omega_{z}\langle z p_z \rangle, \nonumber \\
        & m^2 \omega_{y}^2\langle y^2 \rangle, \:\:
        \langle p_{y}^2 \rangle, \:\:
        m \omega_{y} \langle y p_y \rangle, \nonumber \\
        & m^2 \omega_{x}^2 \langle x^2 \rangle, \:\:
        \langle p_{x}^2 \rangle, \:\:
        m \omega_{x} \langle x p_x \rangle \big]^T. \label{eq:Enskog_state_vector}
\end{align}
The linearized Enskog equations can then be written in the succinct form 
\begin{align}
    \dot{\boldsymbol{\xi}}(t) = \boldsymbol{\Phi}(\Theta) \boldsymbol{\xi}(t), \label{eq:Enskog_linsys}
\end{align}
where the overhead-dot denotes a time derivative and $\boldsymbol{\Phi}$ is a state-relation matrix with units of frequency:

\onecolumngrid
\begin{align}
    \boldsymbol{\Phi}(\Theta) = \begin{bmatrix}
    0 & 0 & 2\omega_{z} & 0 & 0 & 0 & 0 & 0 & 0 \\
    0 & \Phi_{22}(\Theta) & -2 \omega_z & 0 & \Phi_{25}(\Theta) & 0 & 0 & \Phi_{28}(\Theta) & 0 \\
    -\omega_z & \omega_{z} & 0 & 0 & 0 & 0 & 0 & 0 & 0 \\
    0 & 0 & 0 & 0 & 0 & 2\omega_{y} & 0 & 0 & 0 \\
    0 & \Phi_{52}(\Theta) & 0 & 0 & \Phi_{55}(\Theta) & -2\omega_{y} & 0 & \Phi_{58}(\Theta) & 0 \\
    0 & 0 & 0 & -\omega_{y} & \omega_{y} & 0 & 0 & 0 & 0 \\
    0 & 0 & 0 & 0 & 0 & 0 & 0 & 0 & 2\omega_{x} \\
    0 & \Phi_{82}(\Theta) & 0 & 0 & \Phi_{85}(\Theta) & 0 & 0 & \Phi_{88}(\Theta) & -2\omega_{x} \\
    0 & 0 & 0 & 0 & 0 & 0 & -\omega_{x} & \omega_{x} & 0
\end{bmatrix}, \label{eq:Enskog_matrix}
\end{align} 
\twocolumngrid

\noindent
with $\Phi_{ij}(\Theta)$ being the collision integral associated terms that can be read off from Eqs.~(13a--13c) of Ref.~\cite{Wang20_PRA} for bosons, or Eqs.~(\ref{eq:collint_fermions_a}\textcolor{blue}{--}\ref{eq:collint_fermions_c}) of the present paper for fermions. These $\Phi_{ij}$ terms are responsible for coupling the axes, allowing for cross-dimensional rethermalization.  

This first-order linear system of differential equations can be solved via matrix exponentiation. First diagonalize $\boldsymbol{\Phi}$ as
\begin{align}
    \boldsymbol{\Phi} = \boldsymbol{X} \boldsymbol{\Omega} \boldsymbol{X}^{-1},
\end{align}
where $\boldsymbol{X}$ has eigenvectors  of $\Phi$ as columns, and $\boldsymbol{\Omega}$ is a diagonal matrix of 9 complex eigenvalues, $\Omega_k = \Gamma_k + i \omega$. The real parts of each $\Omega_k$ constitute decay rates while imaginary parts, oscillation frequencies. The resulting solution for $\boldsymbol{\xi}$ is then written as
\begin{align}
    \boldsymbol{\xi}(t) = \exp\left(  \boldsymbol{\Phi} t \right) \boldsymbol{\xi}(0) = \boldsymbol{X} \exp\left(  \boldsymbol{\Omega} t \right) \boldsymbol{X}^{-1} \boldsymbol{\xi}(0). \label{eq:Enskog_soln}
\end{align}
The initial value ${\bf \xi}(0)$ represents the specific way in which the gas is brought out of equilibrium, assuming nevertheless that it describes an approximately Gaussian distribution.  Thus, for example, a gas that has equilibrium temperature $T_0$ in the $x$ and $y$ directions, but has temperature ${\cal T}_z$ raised in the $z$ direction at time $t = 0$, would be described by the initial condition
\begin{align}
    \boldsymbol{\xi}(0) = m k_B T_0 \big[ r, r, 0, 1, 1, 0, 1, 1, 0 \big]^T,
\end{align} 
where $r$ denotes the excitation ratio ${\cal T}_z / T_0$. 

Pseudotemperatures defined in Eq.~(\ref{eq:pseudotemperatures}), can be extracted by left multiplication with the row vectors
\begin{subequations}
\begin{align}
    & \boldsymbol{R}_z = \left[ 1, 1, 0, 0, 0, 0, 0, 0, 0 \right] / (2 m k_B), \\
    & \boldsymbol{R}_y = \left[ 0, 0, 0, 1, 1, 0, 0, 0, 0 \right] / (2 m k_B), \\
    & \boldsymbol{R}_x = \left[ 0, 0, 0, 0, 0, 0, 1, 1,  0 \right] / (2 m k_B),
\end{align}
\end{subequations}
to $\boldsymbol{\xi}$, to give $\mathcal{T}_z, \mathcal{T}_y$ and $\mathcal{T}_x$ respectively. The time evolution of the pseudotemperatures defined in this way can be directly compared to that of experimental data.  In such a case, the analytic time trace can be fit to a decaying exponential:
\begin{align}
    T_{\text{fit}}(t) = T_{\text{eq}} + \left[ \mathcal{T}_j(0) - T_{\text{eq}} \right] \exp( - t / \tau_{ij} ), \label{eq:exp_fit}
\end{align}
in the same way the data are. Here $T_{\text{eq}}$ is the equilibration temperature as calculated in App.~\ref{app:NCPR_derivation}. The number of collisions per rethermalization is then computed by taking the reciprocal $\gamma_{ij} = 1 / \tau_{ij}$, and plugging that into Eq.~(\ref{eq:gamma_N_rln}).

Alternatively, one can find the $1/e$ decay time by directly determining the root $\tau_{ij}$, from the relation
\begin{align}
    & \frac{ {\cal T}_j(\tau_{ij}) - T_{\text{eq}} }{ {\cal T}_j(0) - T_{\text{eq}} } = \frac{ \boldsymbol{R}_j \exp( \boldsymbol{\Phi} \tau_{ij} ) \boldsymbol{\xi}_i - T_{\text{eq}} }{ \boldsymbol{R}_j \boldsymbol{\xi}_i - T_{\text{eq}} } = \frac{1}{e}, \label{eq:1e_decay_time} 
\end{align}
based on the pseudotemperature curves from explicit time evolution of the Enskog equations, $\boldsymbol{R}_j \boldsymbol{\xi}(t)$. Above, $\boldsymbol{\xi}_i$ is the initial state $\boldsymbol{\xi}(0)$, with excitation along axis $i$. Such solutions can be obtained directly from Eq.~(\ref{eq:1e_decay_time}) through numerical solvers (e.g. the Newton-Raphson method). This latter procedure is how we determine lifetimes from the full Enskog solutions in what follows. 



\subsection{ \label{sec:short_time_approx} Anisotropic Relaxation: Short-Time Approximation }

The return to equilibrium, in general, is a multi-exponential function of time, owing to the multiple decay rates set by the eigenvalues of $\Phi$. In general, effective relaxation rates are best determined by fitting the exact time evolution.  However, in many cases it may occur that a single rate is dominant, in which case the rate can be determined from the short-time behavior of the decay. This circumstance permits a derivation of analytic expressions for ${\cal N}$. We refer to this scheme as the ``short-time approximation". To formulate this approximation, we define the phase space averaged quantity
\begin{align}
    \langle \chi_j \rangle &\equiv k_B (\mathcal{T}_j - T_{\text{eq}}),
\end{align}
which quantifies the system's deviation from its equilibration temperature, $T_{\text{eq}}$. From Eqs.~(\ref{eq:Enskog_eqns}), the relaxation of $\langle \chi_j \rangle$ would then follow the differential equation
\begin{align}
    \frac{ d \langle \chi_j \rangle }{ dt } = {\cal C}[ \chi_j ] = \frac{\mathcal{C}[ p_j^2 ]}{2 m}.
\end{align}
We now assert that for small deviations from equilibrium and at short times, this can be approximated with a decay rate $\gamma$, as ${\cal C}[ \chi_j] \approx - \gamma \langle \chi_j \rangle$, which results in the relation
\begin{align}
    \gamma = - \left. \frac{1}{ \left( \mathcal{T}_j(t) - T_{\text{eq}} \right) } \frac{ d \mathcal{T}_j(t) }{dt} \right|_{t = 0}. \label{eq:retherm_rate_theory}
\end{align}

Using this $\gamma$ and the standard collision rate $n {\bar \sigma} \langle v \rangle$, we can extract the value of ${\cal N}_{ij}$ via the relation in Eq.~(\ref{eq:gamma_N_rln}). 
As before, we consider an excitation of axis $i$, following which the rethermalization rate is measured along axis $j$. This is modeled by taking axis $i$ to have an initial out-of-equilibrium pseudotemperature
\begin{align}
    \mathcal{T}_i = T_{0} + \frac{\delta_i}{k_B},
\end{align}
where $\delta_i$ is a perturbance to the energy, while the initial temperatures along the 2 other axes are simply $T_{0}$. By construction of Eq.~(\ref{eq:retherm_rate_theory}), $\delta_i$ stands as an auxiliary variable which cancels out in the derivation.

\section{\label{sec:NCPR_bosons} Rethermalization in Dipolar Bose Gases}


In ultracold but thermal dipolar Bose gases, analytic expressions for the collision integrals have been derived in Ref.~\cite{Wang20_PRA} within the Enskog formalism. This completes the framework of Eq.~(\ref{eq:Enskog_linsys}), to efficiently extract rethermalization information of the gas with the techniques just discussed. To ground further discussions, we consider a cross-dimensional rethermalization experiment akin to Erbium experiments by the Ferlaino group in Ref.~\cite{Aikawa14_PRL}, using parameters listed in Tab.~\ref{tab:system_parameter} and $a = 39.8$ ($a_0$). 

\begin{table}[H]
\caption{\label{tab:system_parameter} 
Table of parameter values utilized in the Monte Carlo simulation for bosonic $^{166}$Er. Da $= 1.661 \times 10^{-27}$ kg stands for Dalton (atomic mass unit), $a_{0} = 5.292 \times 10^{-11}$ m is the Bohr radius and $\mu_B = 9.274 \times 10^{-24}$ J/T is the Bohr magneton. }
\begin{ruledtabular}
\begin{tabular}{l c c c}
    \multicolumn{1}{c}{\textrm{Parameter}} & \multicolumn{1}{c}{\textrm{Symbol}} & \multicolumn{1}{c}{\textrm{Value}} & \multicolumn{1}{c}{\textrm{Unit}} \\
    \colrule
    Number of particles, & $N$ & 100, 000 & -- \\
    Atomic mass number, & A & 166 & Da \\
    Magnetic moment & $\mu$ & 7 & ($\mu_B$) \\
    Dipole length, & $a_d$ & 99 & ($a_{0}$) \\
    Initial gas temperature, & $T_0$ & 300 & nK \\
    Axial trapping frequency, & $\omega_{z}$ & $2\pi\times 30$ & Hz \\
    Radial trapping frequency, & $\omega_{\perp}$ & $2\pi\times 300$ & Hz
\end{tabular}
\end{ruledtabular}
\end{table}

\onecolumngrid

\begin{figure}[ht]
    \centering
    \includegraphics[width=\columnwidth]{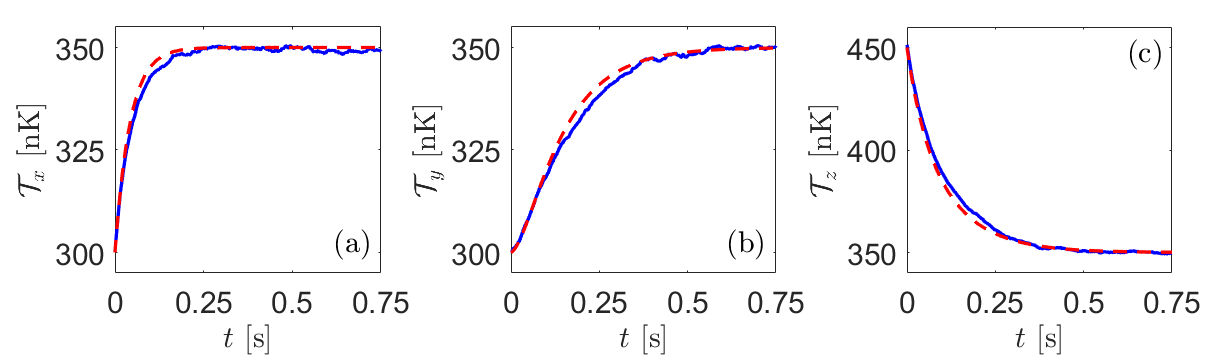}
    \caption{ Rethermalizaton curves of the pseudotemperatures (a) $\mathcal{T}_x$, (b) $\mathcal{T}_y$ and (c) $\mathcal{T}_z$, with excitation along $z$, as generated by DSMC simulations (solid, blue) and solutions to the Enskog equations (dashed, red). Simulation parameters are those in Tab.~\ref{tab:system_parameter}, with the scattering length is set to $a = 39.8$ ($a_0$) and $\Theta = 90$ (deg). }
    \label{fig:DSMCvsEnskog_T_a40_Theta90}
\end{figure}

\twocolumngrid

We assess the validity of the resulting full Enskog solutions by first comparing them to results of DSMC simulations. Such a comparison of the pseudotemperature time traces is shown in Fig.~\ref{fig:DSMCvsEnskog_T_a40_Theta90}. For this example, we chose an initial temperature excitation along the $z$ axis by a factor $\mathcal{T}_z(0) = 1.5 T_0$. This allows us to observe the collective rethermalization behavior above the Monte Carlo stochastic noise. The DSMC result is shown in blue (solid curves), while the result from the full Enskog solution is in red (dashed curves), where the dipolar axis is tilted at an angle $\Theta = 90$ (deg).

\begin{figure}[ht]
    \centering
    \includegraphics[width=\columnwidth]{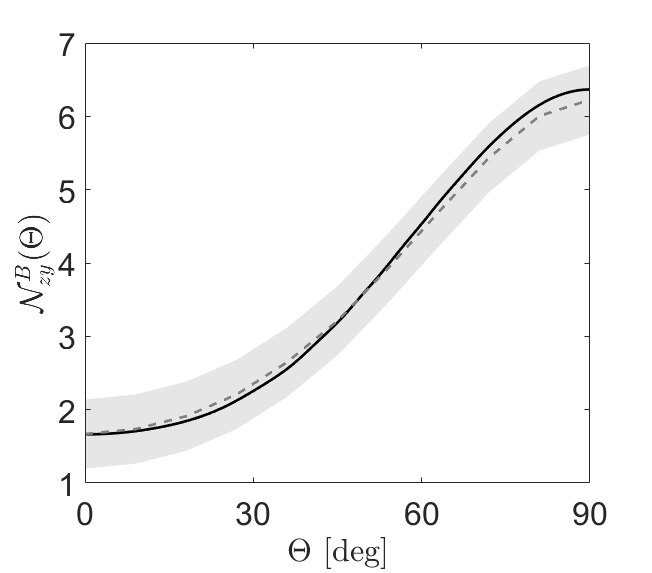}
    \caption{ The number of collisions per rethermalization ${\cal N}^B_{zy}$ as a function of $\Theta = 0$--$90$ (deg), obtained from the theory [Eq.~(\ref{eq:1e_decay_time})] (solid, black curve) and fits to ${\cal T}_y$ from DSMC simulations using Eq.~(\ref{eq:exp_fit}) (dashed-dotted, grey data) with error bars (filled, grey region).  These results are with the parameters in Tab.~\ref{tab:system_parameter}, and $a = 39.8$ ($a_0$).}
    \label{fig:NCPRzy_FitVsTheory}
\end{figure}

For all three pseudotemperatures the agreement of the two methods is excellent, thus justifying the use of the Enskog equations for extracting ${\cal N}$. As an added measure of surety, we also compare ${\cal N}$ obtained from the full Enskog solution of Eq.~(\ref{eq:1e_decay_time}), to fits to the DSMC data (with error region obtained through fits to repeated Monte Carlo trials) at $\Theta = 0$ -- $90$ (deg). These too, show agreement as seen in Fig.~\ref{fig:NCPRzy_FitVsTheory}. On top of this, our theoretical result reverts to ${\cal N} = {\cal N}_s = 2.5$ in the absence of dipoles, $\mu = 0$. Thus, solutions obtained from the Enskog equations provide a simple alternative to the full mathematical machinery of the DSMC method in modeling relaxation.

We next consider the alternative method proposed to compute the number of collisions per rethermalization from the Enskog equations, by using the short-time approximation (Sec.~\ref{sec:short_time_approx}). This method, while far simpler, is not always as accurate, as we will see. 

For clarity of presentation, we define an effective isotropic scattering length \cite{Wang20_PRA} and length scale:
\begin{subequations}
\begin{align}
    & a_{\text{eff}}^2 \equiv 2\left( a^2 - \frac{ 4 }{ 3 } a a_d + \frac{ 4 }{ 9 } a_d^2 \right), \\
    & \ell^2 \equiv 4 \left( a^2 + \frac{ 4 }{ 45 } a_d^2 \right),
\end{align}
\end{subequations}
respectively. Moreover, the results are usefully presented in terms of the dimensionless length quantities $\tilde{a}_{\text{eff}} = a_{\text{eff}} / a$ and $\tilde{a}_d = a_d / a$, scaled in units of the scattering length. 

The resulting analytical expressions for ${\cal N}_{ij}$ possess intricate denominators, whereby it is useful to write the expressions in the form
\begin{align}
    {\cal N}^B_{ij}(\Theta) \equiv \frac{(\ell / a)^2}{\Lambda^B_{ij}}.
\end{align}

This permits the functional forms: 
\begin{widetext}
\begin{subequations}
\label{eq:NCPR_bosons}
\begin{align}
    & \Lambda^B_{xx}(\Theta) = { \tilde{a}_{\text{eff}}^2 \left( \frac{4}{5} \right) + \tilde{a}_d \left( \frac{16}{35} \right) \left[ 5 - \cos (2 \Theta ) \right] + \tilde{a}_d^2 \left( \frac{1}{105} \right) \left[ \cos (4 \Theta ) + 4 \cos (2 \Theta ) - 61 \right]}, \label{eq:NCPR_bosons_a} \\
    & \Lambda^B_{yx}(\Theta) = { \tilde{a}_{\text{eff}}^2 \left( \frac{4}{5} \right) + \tilde{a}_d \left( \frac{32}{35} \right) \left[ 2 - \cos (2 \Theta ) \right] + \tilde{a}_d^2 \left( \frac{8}{105} \right) \left[ \cos (2 \Theta ) - 7 \right]}, \label{eq:NCPR_bosons_b} \\
    & \Lambda^B_{zx}(\Theta) = { \tilde{a}_{\text{eff}}^2 \left( \frac{4}{5} \right) + \tilde{a}_d \left( \frac{96}{35} \right) + \tilde{a}_d^2 \left( \frac{2}{105} \right) \left[ \cos(4 \Theta) - 33 \right]}, \label{eq:NCPR_bosons_c} \\
    & \Lambda^B_{yy}(\Theta) = { \tilde{a}_{\text{eff}}^2 \left( \frac{4}{5} \right) + \tilde{a}_d \left( \frac{64}{35} \right) - \tilde{a}_d^2 \left( \frac{56}{105} \right)}, \label{eq:NCPR_bosons_d} \\
    & \Lambda^B_{zy}(\Theta) = { \tilde{a}_{\text{eff}}^2 \left( \frac{4}{5} \right) + \tilde{a}_d \left( \frac{32}{35} \right) \left[ 2 + \cos(2\Theta) \right] - \tilde{a}_d^2 \left( \frac{8}{105} \right) \left[ \cos(2\Theta) + 7 \right]}, \label{eq:NCPR_bosons_e} \\
    & \Lambda^B_{zz}(\Theta) = { \tilde{a}_{\text{eff}}^2 \left( \frac{4}{5} \right) + \tilde{a}_d \left( \frac{16}{35} \right) \left[ 5 + \cos (2 \Theta ) \right] + \tilde{a}_d^2 \left( \frac{1}{105} \right) \left[ \cos(4\Theta) - 4 \cos (2 \Theta ) - 61 \right]}. \label{eq:NCPR_bosons_f}
\end{align}
\end{subequations}
\end{widetext}

Several configurations are omitted due to the symmetries $\Lambda^B_{yx}(\Theta) = \Lambda^B_{xy}(\Theta)$, $\Lambda^B_{zy}(\Theta) = \Lambda^B_{yz}(\Theta)$ and $\Lambda^B_{zx}(\Theta) = \Lambda^B_{xz}(\Theta)$. These arise because $\Theta$ is defined in the $x,z$-plane, resulting in a $\pi/2$ periodicity about the $y$-axis.  A detailed derivation of these expressions is provided in Appendix~\ref{app:NCPR_derivation}. These formulas once again obtain ${\cal N} = 2.5$ with pure $s$-wave scattering.

A striking result is that these expressions do not depend on the 3-dimensional trap shape. They are thus robust to arbitrary trap geometries in the perturbative limit, provided the trap remains harmonic. For bosons, the effective variables of interest are then the dipole alignment angle $\Theta$, and reduced dipole length $\tilde{a}_d$. The latter  can be varied in an experiment by either changing the dipole moment (e.g. through varying the strength of the electric field applied to polar molecules), or changing the scattering length via the multitude of Fano-Feshbach resonances in the lanthanide series. 

\begin{figure}[ht]
    \centering
    \includegraphics[width=\columnwidth]{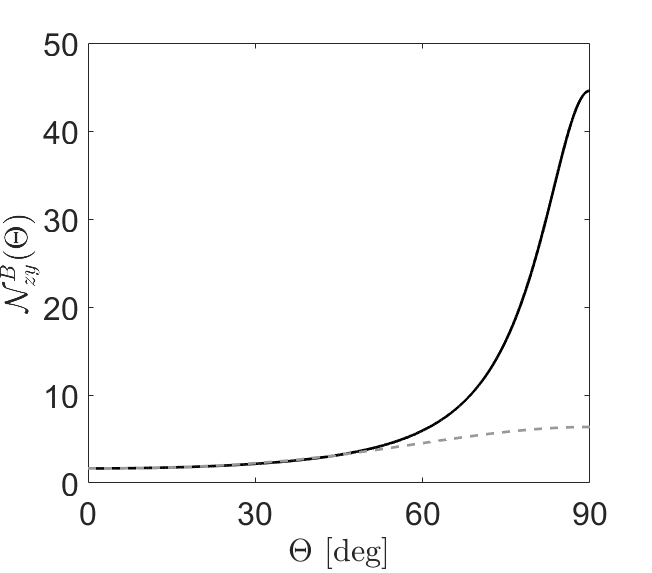}
    \caption{ The number of collisions per rethermalization, ${\cal N}^B_{zy}$ vs $\Theta = 0$--$180$ (deg), as computed using the short-time approximation (solid, black curve) and the full Enskog solution (dashed, gray curve). These results are with the parameters in Tab.~\ref{tab:system_parameter}, and $a = 39.8$ ($a_0$). }
    \label{fig:NCPRzy_FEvsSTA}
\end{figure}

The analytic expressions of Eqs.~(\ref{eq:NCPR_bosons}) are comparatively concise, making them appealing for application to experiments. They do however, come with caveats which arise when the assumption of a single decay rate ceases to hold (i.e. clear non-exponential behavior is present). This can be seen, for instance, by comparing the short-time result to the number of collisions per rethermalization in the example considered above.  This comparison is made in Fig.~\ref{fig:NCPRzy_FEvsSTA}. Here the dashed curve is the full Enskog result, reproducing the solid curve in Fig.~\ref{fig:NCPRzy_FitVsTheory}; while the solid line is the short-time approximation from Eq.~(\ref{eq:NCPR_bosons_e}). This figure shows that the short-time approximation expression works well at small angles, while it greatly overestimates ${\cal N}_{zy}$ near $\Theta = 90$ (deg). The Enskog solution was already shown in Fig.~\ref{fig:DSMCvsEnskog_T_a40_Theta90}, so it is clear that the fault stems from taking the short-time limit.

\begin{figure}[ht]
    \centering
    \includegraphics[width=\columnwidth]{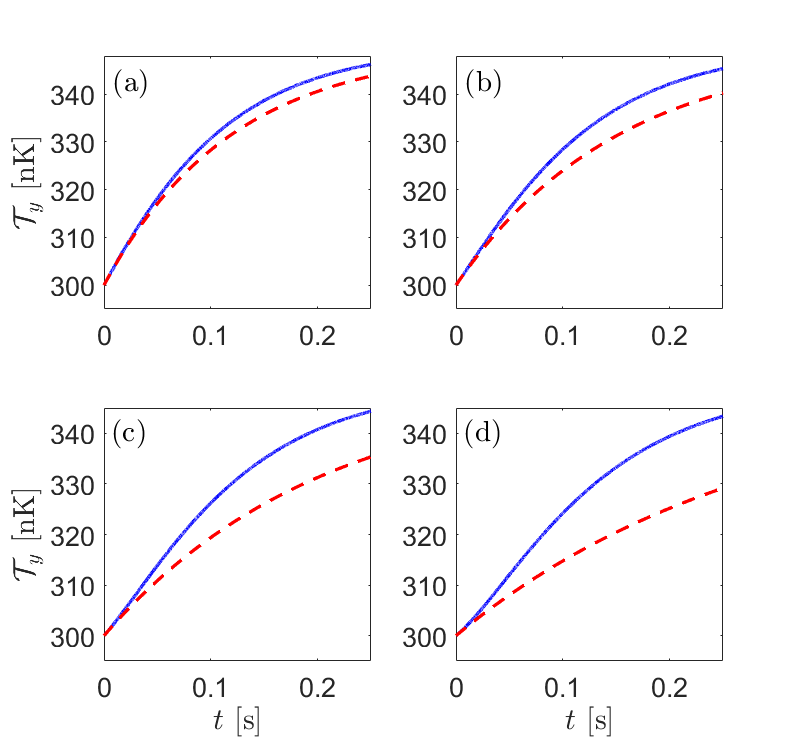}
    \caption{ Rethermalization curves ${\cal T}_y(t)$ obtained from the Enskog equation (solid, blue); compared to exponential relaxation curves with decay rates obtained from the analytic formulas in Eqs.~(\ref{eq:NCPR_bosons}) (dashed, red). The red curves are appropriate at $t=0$, hence do not model the S-shape character of the Enskog solutions, which is more pronounced as $\Theta$ is increased, illustrated in subplots: (a) $\Theta = 55$ (deg), (b) $\Theta = 60$ (deg), (c) $\Theta = 65$ (deg) and (d) $\Theta = 70$ (deg). }
    \label{fig:T_a40_Semerge}
\end{figure}

The issue with the short-time approximation can be seen in more detailed time traces, in Fig.~\ref{fig:T_a40_Semerge}, for four different angles $\Theta$.  In each case, the Enskog curve (solid, blue) shows an initial  ``S"-shaped curvature, thus a fit to short times yields an incorrect slope and the wrong exponential decay, as shown by the red, dashed curves.

\begin{figure}[ht]
    \centering
    \includegraphics[width=\columnwidth]{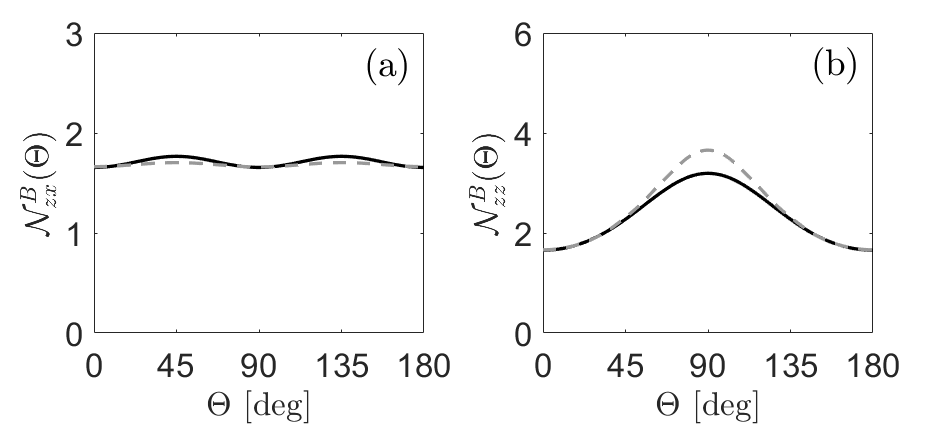}
    \caption{ The number of collisions per rethermalization (a) ${\cal N}^B_{zx}$ and (b) ${\cal N}^B_{zz}$ vs $\Theta = 0$--$180$ (deg), as computed using the short-time approximation (solid, black curve) and the full Enskog solution (dashed, grey curve). These results are with the parameters in Tab.~\ref{tab:system_parameter}, and $a = 39.8$ ($a_0$). }
    \label{fig:NCPRzx_zz_FEvsSTA}
\end{figure}

We have deliberately chosen these examples to emphasize the possible discrepancy between the full Enskog solution, which is demonstrably quite accurate, and the short-time approximation, which is still often accurate. To illustrate the accuracy of the latter, we present in Fig.~\ref{fig:NCPRzx_zz_FEvsSTA} results for the same experiment described above but with rethermalization measured along the $x$ and $z$ axes, rather than the $y$ axis. This is plot up to $\Theta = 180$ (deg) instead of $\Theta = 90$ (deg), showing a much closer quantitative correspondence between the full Enskog solution and short-time approximation. As a practical matter, therefore, we recommend using the analytical formulas (\ref{eq:NCPR_bosons}) where they adequately describe data, but to defer to the more complete numerical solutions of Eq.~(\ref{eq:Enskog_soln}) when necessary. 

The expressions for ${\cal N}$ point to intriguing possibilities.  For example, it is notable that ${\cal N}^B_{zx}$ remains consistently below $2.5$ (pure $s$-wave scattering) for all values of $\Theta$. This opens the possibility for increased efficiency in evaporative cooling, where for instance atoms can be evaporated along the $y$-axis, forcing rethermalization to occur primarily in the directions of preferential scattering. Comprehensive studies of this process would however necessitate a theory which can model the transition into quantum degeneracy. This could, for instance, be done using the Uehling-Uhlenbeck equation \cite{Uehling33_PR, Lepers10_PRA}, or c-field techniques \cite{Blakie08_AP}. We leave further discussions of such optimal evaporative cooling schemes to future works.

\section{ \label{sec:NCPR_fermions} Rethermalization in Dipolar Fermi Gases }

The same analysis can also be performed for ultracold fermionic dipoles, by using the appropriate differential cross section from Ref.~\cite{Bohn14_PRA} in the collision integrals. We derive the linearized collision integrals for fermions in App.~\ref{app:fermi_collision_integrals}. This then allows us to compute $\tau_{ij}$, with both the full Enskog equations and the short-time approximation. 

\begin{figure}[ht]
    \centering
    \includegraphics[width=\columnwidth]{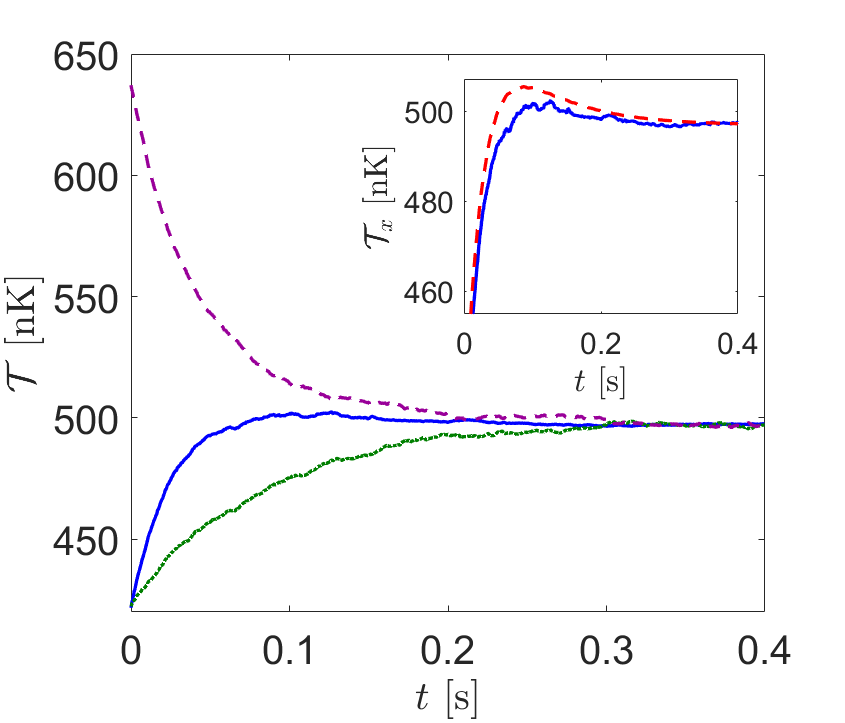}
    \caption{ Rethermalizaton curves of the pseudotemperatures $\mathcal{T}_x$ (solid, blue), $\mathcal{T}_y$ (dotted, green) and $\mathcal{T}_z$ (dashed, purple), with excitation along $z$. Simulation parameters are those in Tab.~\ref{tab:system_parameter2}, with $\Theta = 45$ (deg). The inset zooms in on the nonmonotic hump of $\mathcal{T}_x$, comparing the Monte Carlo solution (solid, blue) with the Enskog solution (dashed, red). }
    \label{fig:T_F_Theta45}
\end{figure}

As for bosons, we first show that the full Enskog solution is robust by comparing it to DSMC simulations. We do so by once again modeling an experiment where non-exponential behavior is prominent, namely that described in Ref.~\cite{Sykes15_PRA} (parameters listed in Tab.~\ref{tab:system_parameter2}) with fermionic $^{167}$Er atoms \footnote{ We have redefined the coordinate axes from those in Ref.~\cite{Sykes15_PRA}, so as to make $z$ the axis of cylindrical symmetry. }. In this case the initial excitation is set to $\mathcal{T}_z(0) = 1.5 T_0$.  The resulting rethermalization is shown in the time-dependent pseudotemperatures in Fig.~\ref{fig:T_F_Theta45}, for $\Theta = 45$ (deg).

\begin{table}[ht]
\caption{\label{tab:system_parameter2} 
Table of parameter values utilized in the Monte Carlo simulation for fermionic $^{167}$Er. }
\begin{ruledtabular}
\begin{tabular}{l c c c}
    \multicolumn{1}{c}{\textrm{Parameter}} & \multicolumn{1}{c}{\textrm{Symbol}} & \multicolumn{1}{c}{\textrm{Value}} & \multicolumn{1}{c}{\textrm{Unit}} \\
    \colrule
    Number of particles, & $N$ & 80, 000 & -- \\
    Atomic mass number, & A & 167 & Da \\
    Magnetic moment & $\mu$ & 7 & ($\mu_B$) \\
    Dipole length, & $a_d$ & 99 & ($a_{0}$) \\
    Initial gas temperature, & $T_0$ & 456 & nK \\
    Axial trapping frequency, & $\omega_{z}$ & $2\pi\times 40$ & Hz \\
    Radial trapping frequency, & $\omega_{\perp}$ & $2\pi\times 400$ & Hz 
\end{tabular}
\end{ruledtabular}
\end{table}
 
\begin{figure}[ht]
    \centering
    \includegraphics[width=\columnwidth]{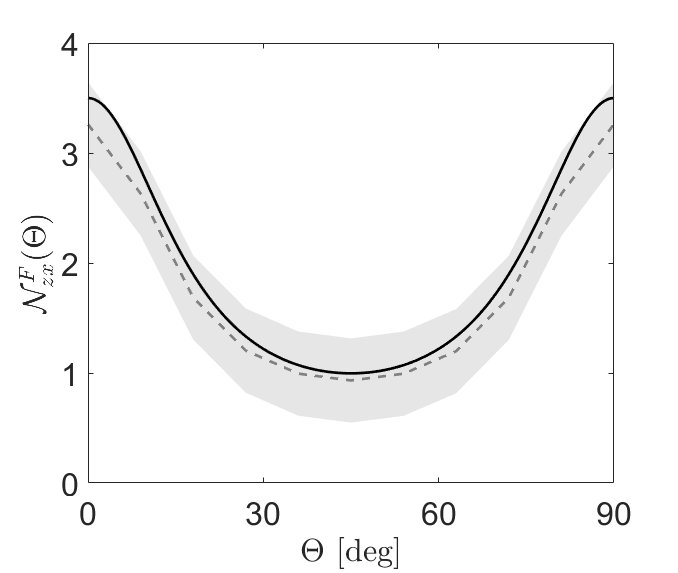}
    \caption{ The number of collisions per rethermalization ${\cal N}^F_{zx}$ as a function of $\Theta = 0$--$90$ (deg), obtained from the theory [Eq.~(\ref{eq:1e_decay_time})] (solid, black curve) and exponential fits to ${\cal T}_x$ from DSMC simulations using Eq.~(\ref{eq:exp_fit}) (dashed-dotted, grey data) with error bars (filled, grey region). }
    \label{fig:NCPRzx_F_FitVsTheory}
\end{figure}

In this particular example, the relaxation of ${\cal T}_z$ is nonexponential, as seen in the nonmonotonic behavior shown in the inset.  This is reminiscent of the behavior of an almost-overdamped oscillator. This behavior is captured by the Enskog solutions (red dashed lines), leading again to a reasonable description of the rethermalization.  Because of this, the number of collisions per rethermalization extracted from the Enskog equations still shows good agreement with that obtained through exponential fits to Monte Carlo simulations as presented in Fig~\ref{fig:NCPRzx_F_FitVsTheory}.

As for the short-time approximation, the absence of $s$-wave scattering in dipolar fermions leaves only a single relevant length scale, $a_d$. The value of $a_d$ is however irrelevant to the value of ${\cal N}$, as ${\cal N}$ represents a ratio resulting in its cancellation. The number of collisions per rethermalization for fermions, ${\cal N}^F_{ij}$, is then a function of $\Theta$ only. The resulting analytical expressions for ${\cal N}^F_{ij}(\Theta)$ take the relatively simple forms:
\begin{subequations} \label{eq:NCPR_fermions}
\begin{align}
    & {\cal N}^F_{xx}(\Theta) = \frac{112}{45 - 4 \cos(2\Theta) - 17 \cos(4\Theta)}, \\
    & {\cal N}^F_{yx}(\Theta) = \frac{14}{3 - \cos(2\Theta)}, \\
    & {\cal N}^F_{zx}(\Theta) = \frac{56}{33 - 17 \cos (4 \Theta )}, \\
    & {\cal N}^F_{yy}(\Theta) = \frac{14}{3}, \\
    & {\cal N}^F_{zy}(\Theta) = \frac{14}{3 + \cos(2\Theta)}, \\
    & {\cal N}^F_{zz}(\Theta) = \frac{112}{45 + 4 \cos(2\Theta) - 17 \cos(4\Theta)},
\end{align}
\end{subequations}
once again with omissions since ${\cal N}^F_{yx}(\Theta) = {\cal N}^F_{xy}(\Theta)$, ${\cal N}^F_{zy}(\Theta) = {\cal N}^F_{yz}(\Theta)$ and ${\cal N}^F_{zx}(\Theta) = {\cal N}^F_{xz}(\Theta)$, due to the experimentally imposed symmetries.

\begin{figure}[ht]
    \centering
    \includegraphics[width=\columnwidth]{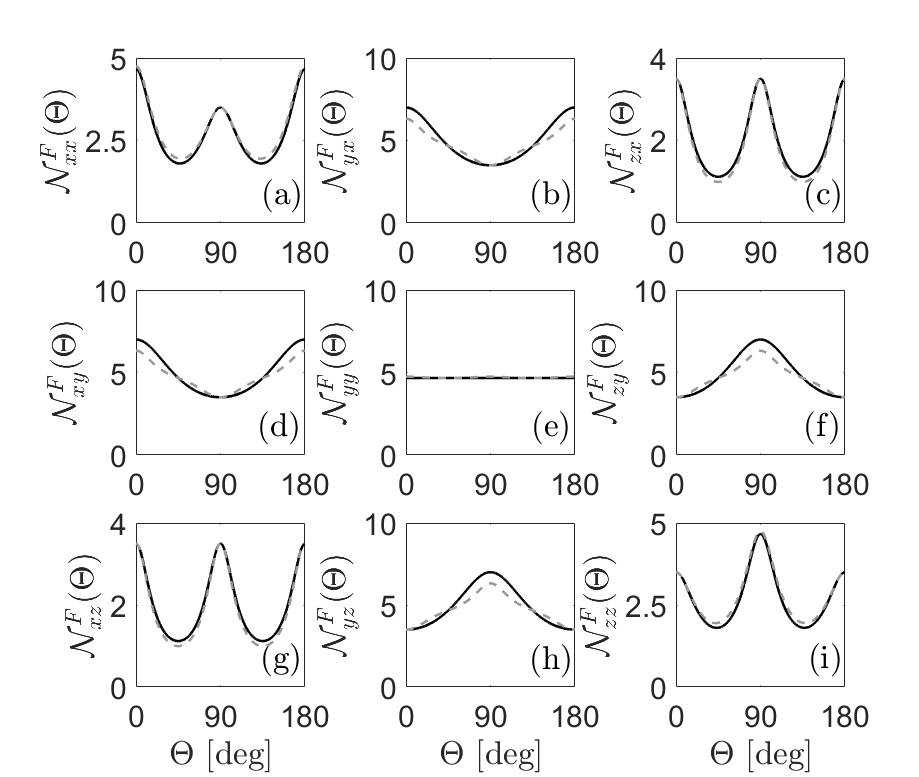}
    \caption{ The number of collisions per rethermalization for fermions ${\cal N}^F(\Theta)$ vs $\Theta = 0$--$180$ (deg), for all nine excitation-rethermalization configurations from the short-time approximation (solid, black curves) and full Enskog solution (dashed, grey curves). }
    \label{fig:NCPR_fermions_2def}
\end{figure}

Having ${\cal N}$ as a single-variable function of $\Theta$ permits a comprehensive visualization of its behavior with ease. We provide the plots of ${\cal N}^F$, derived with both the full Enskog solutions (dashed, grey curves) and short-time approximation (solid, black curves) in Fig.~\ref{fig:NCPR_fermions_2def}. We see that in the case of fermions, both these definitions result in very similar values for the number of collisions per rethermalization. Thus in contrast to the boson case, for fermions the simple formulas of the short-time approximation should always be useful in describing rethermalization data.

\begin{figure}[ht]
    \centering
    \includegraphics[width=\columnwidth]{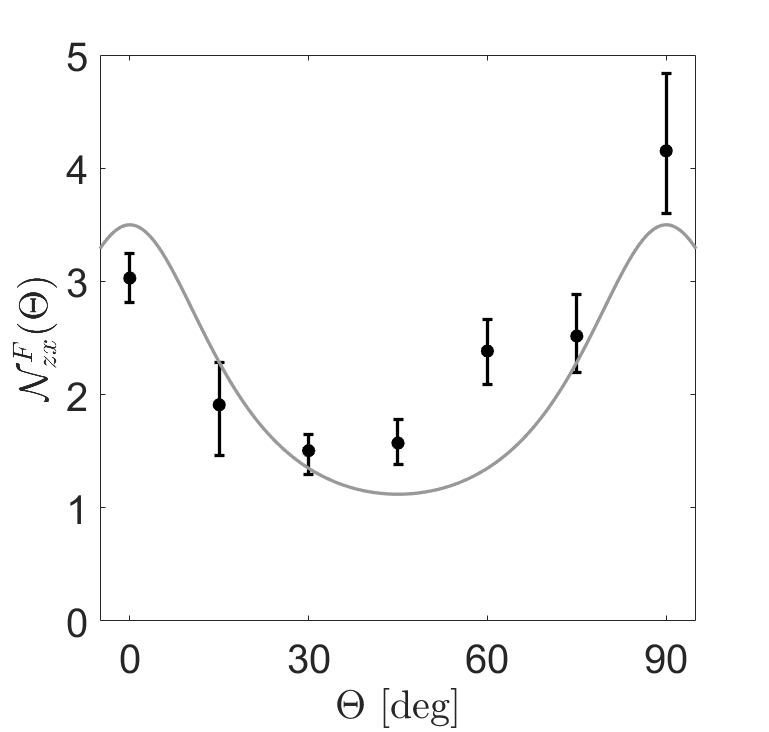}
    \caption{ The number of collisions per thermalization ${\cal N}^F_{zx}(\Theta)$. The solid (grey) curve shows the theoretical results from Eq.~(\ref{eq:NCPR_fermions}), whereas the (black) circles with error bars are results from the KRb experiment at JILA. }
    \label{fig:NCPRzx_F_ExpVsTheory}
\end{figure}

As an example, we show in Fig.~\ref{fig:NCPRzx_F_ExpVsTheory} a comparison of the formula for ${\cal N}_{z,x}^F$ to the same quantity extracted from  recent rethermalization data for KRb molecules in an experiment at JILA \cite{Li21_ArXiv}. Notably, the formula describes the anisotropy as is, without any adjustable parameters. This  result illustrates the efficacy  of the theory for ultracold polar molecules, beyond its originally intended application to magnetic atoms.

\onecolumngrid

\twocolumngrid

An unusual circumstance in Ref.~\cite{Li21_ArXiv} is that the experiment is performed at a high electric field ${\cal E} = 12.72$ kV/cm, where the intermoleclar potential energy surface is distorted into a shielding configuration that prevents the molecules from reacting chemically.  An analysis of this surface \footnote{L. Lassabli\`ere, and G.  Qu\'em\'ener  (private communication).} reveals that, near this field, the interaction between the two molecules is nearly equivalent to the usual dipole-dipole interaction, but with reversed sign. In this case the differential scattering cross section, as evaluated in the Born approximation, is equivalent to that for the ordinary scattering of dipoles, from which the same  formulas as in Eqs.~(\ref{eq:NCPR_fermions}) follow.

\section{ \label{sec:conclusion} Conclusions }

En route to equilibrium, dipolar anisotropies manifest in the thermalization rate variations between axes of trapped ultracold but thermal gases. This is quantified by the so-called number of collision per rethermalization ${\cal N}$, as a function of dipole-alignment angle $\Theta$, which also serves as a measure of efficiency to which collisions count toward thermalization. In close-to-equilibrium scenarios, the linearized Enskog equations allow convenient computation of ${\cal N}$, following two theoretical procedures: a) the full Enskog solution; b) the short-time approximation. 

The latter of these (b), results in surprisingly simple yet accurate (for fermions and many instances of bosons) analytic expressions, that utilize only short-time rethermalization information. These expressions do however have limits to their use, elucidated through the example rethermalization simulation of Tab.~\ref{tab:system_parameter}. Conversely, we showed that the former method (a), an operational $1/e$ decay construction of ${\cal N}$, more closely resembles what one would extract from experimental rethermalization data. This however is more involved than its analytic counterpart, but is still obtained with much greater ease than through numerous trials of explicit numerical particle simulations (e.g. via Monte Carlo methods). 

Our prescription is thus to utilize the analytic formulas (b), for systems of ultracold fermions, and preliminary studies of ultracold bosons. Caution should however be taken with bosons, especially if large suppression/enhancement of rethermalization is predicted. In such cases, we advocate the use of the full Enskog solution (a), to provide a more accurate representation of what to expect from experiments.   

We also point out that the inherent anisotropy of ${\cal N}$, presents the opportunity for engineering enhanced evaporative cooling trajectories. The current models of this work are however inadequate around the transition temperatures $T_c$, for bosons and $T_F$, for fermions. We defer treatments of such issues and optimal evaporation protocols to a future publication.


\onecolumngrid

\begin{acknowledgments}

This material is based upon work supported by the National Science Foundation under Grant Number PHY 1734006 and PHY 1806971. We thank the JILA KRb group for sending us their experimental data, and for engaging discussions.

\end{acknowledgments}

\appendix

\section{ \label{app:NCPR_derivation} Rethermalization in the Short-Time Approximation }

An explicit derivation of the number of collisions per rethermalization ${\cal N}^B(\Theta)$, for configuration $(i, j) = (x, x)$, in the short-time approximation is presented here. This derivation will be instructive for constructing all other ${\cal N}_{ij}(\Theta)$ quantities with a similar procedure. In this particular case, having the excitation along the $x$ axis results in the equilibration temperature
\begin{align}
    T_{\text{eq}} &= \frac{2}{3} T_0 + \frac{1}{3} \left( T_0 + \frac{\delta_x}{k_B} \right) = T_0 + \frac{\delta_x}{3 k_B},
\end{align}
which follows from the equipartition theorem. To now measure the rethermalization along the $x$-axis,  the initial pseudotemperature deviation is inserted into the collision integral $\mathcal{C}[ p_x^2 ]$ of Ref.~\cite{Wang20_PRA}, to give 
\begin{align}
    \mathcal{C}[ p_x^2 ] (\delta_x) &= \left( \dfrac{8 AN a_{\text{eff}}^2}{15 \pi k_B} \right) \left[ -2\delta_x \right] + a_d \left(\frac{64 AN a}{105\pi k_B}\right) \Big[ \delta_x \cos (2 \Theta ) - 5 \delta_x \Big] \nonumber \\
    &\quad + a_d^2 \left(\frac{4 AN}{315\pi k_B}\right)  \Big[ 61 \delta_x - 4\delta_x \cos (2 \Theta ) - \delta_x \cos (4 \Theta ) \Big], 
\end{align}
where $A \equiv m^2 \overline{\omega}^3 / T_0$ and $\overline{\omega}$ is the geometric mean of trapping frequencies. Dividing this by $2m$ gives $d \langle \chi_x \rangle / dt$, which can be used to derive $\gamma$ as
\begin{align}
    \gamma &= -\frac{1}{\langle \chi_x \rangle} \frac{d\langle \chi_x \rangle}{dt} (\delta_x) \nonumber \\
    &= a_{\text{eff}}^2 \left( \dfrac{4 A N}{5 \pi m k_B} \right) + a_d a \left(\frac{16 A N}{35 \pi m k_B} \right) [ 5 - \cos (2 \Theta ) ] + a_d^2 \left( \frac{A N}{105 \pi m k_B} \right) [ \cos (4 \Theta ) + 4 \cos (2 \Theta ) - 61 ].
\end{align}
Finally to get ${\cal N}$, we compute the mean collision rate as
\begin{align}
    n \overline{\sigma} \langle v \rangle = \frac{N}{\pi} \frac{A \ell^2}{k_B}, 
\end{align}
which we divide by $\gamma$ to give
\begin{align}
    {\cal N}^B_{xx}(\Theta) = \frac{4 \left( a^2 + \frac{4}{45} a_d^2 \right)}{ a_{\text{eff}}^2 \left( 4/5 \right) + a_d a \left(16/35\right) \left[ 5 - \cos (2 \Theta ) \right] + a_d^2 \left(1/105\right) \left[ \cos (4 \Theta ) + 4 \cos (2 \Theta ) - 61 \right]}.
\end{align}

\section{\label{app:fermi_collision_integrals} Fermionic Collision Integrals}

By using the prescription provided in App.~A of Ref.~\cite{Wang20_PRA}, we derive the linearized collision integrals for fermions within the Enskog formalism as:
\begin{subequations} \label{eq:collint_fermions}
\begin{align}
\begin{split} \label{eq:collint_fermions_a}
    \mathcal{C}^{F}[ p_x^2 ] &= \left( \frac{4 N }{315 \pi} \right) \left( \frac{a_d^2 m \overline{\omega}^3 }{k_B T_0} \right) \big[ 4 \left( \langle p_x^2 \rangle - \langle p_y^2 \rangle \right) \cos (2 \Theta ) + 17 \left( \langle p_x^2 \rangle - \langle p_z^2 \rangle \right) \cos (4 \Theta ) \\
    &\quad\quad\quad\quad\quad\quad\quad\quad\quad\quad - 45 \langle p_x^2 \rangle + 12 \langle p_y^2 \rangle + 33 \langle p_z^2 \rangle \big],
\end{split} \\
\begin{split} \label{eq:collint_fermions_b}
    \mathcal{C}^{F}[ p_y^2 ] &= \left( \frac{16 N }{315 \pi} \right) \left( \frac{a_d^2 m \overline{\omega}^3 }{k_B T_0} \right) \left[ 3 \langle p_x^2 \rangle - 6 \langle p_y^2 \rangle + 3 \langle p_z^2 \rangle - \left( \langle p_x^2 \rangle - \langle p_z^2 \rangle \right) \cos (2 \Theta ) \right],
\end{split} \\
\begin{split} \label{eq:collint_fermions_c}
    \mathcal{C}^{F}[ p_z^2 ] &= \left( \frac{4 N }{315 \pi} \right) \left( \frac{a_d^2 m \overline{\omega}^3 }{k_B T_0} \right) \big[ 4 \left( \langle p_y^2 \rangle - \langle p_z^2 \rangle \right) \cos (2 \Theta ) - 17 \left( \langle p_x^2 \rangle - \langle p_z^2 \rangle \right) \cos (4 \Theta ) \\
    &\quad\quad\quad\quad\quad\quad\quad\quad\quad\quad + 33 \langle p_x^2 \rangle + 12 \langle p_y^2 \rangle - 45 \langle p_z^2 \rangle \big].
\end{split}
\end{align}
\end{subequations}

\twocolumngrid

\nocite{*}

\bibliography{main.bib} 

\end{document}